\documentclass[11pt]{article}
\usepackage{amsmath,amssymb,color,graphics,epsfig}
\usepackage[compress]{cite}
\usepackage{hyperref}

\usepackage{amsfonts}
\newcommand{\be}{\begin{equation}}
\newcommand{\ee}{\end{equation}}
\newcommand{\bea}{\setlength\arraycolsep{2pt} \begin{eqnarray}}
\newcommand{\eea}{\end{eqnarray}}

\def\ft#1#2{{\textstyle{\frac{\scriptstyle #1}{\scriptstyle #2} } }}

\def\0{{\sst{(0)}}}
\def\1{{\sst{(1)}}}
\def\2{{\sst{(2)}}}
\def\3{{\sst{(3)}}}
\def\4{{\sst{(4)}}}
\def\5{{\sst{(5)}}}
\def\6{{\sst{(6)}}}
\def\7{{\sst{(7)}}}
\def\8{{\sst{(8)}}}
\def\sst#1{{\scriptscriptstyle #1}}

\begin{document}

	%\begin{flushright}
	%\hfill{MI-TH-1533}
	%\end{flushright}
	
	%\vspace{25pt}
\begin{center}
{\Large {\bf Identifying \textit{doppelg\"ange} Black Holes through Shadow Images}}
		
\vspace{40pt}

Yukun Xu,
Hyat Huang\footnote{hyat@mail.bnu.edu.cn;}, Meng-Yun Lai\footnote{mengyunlai@jxnu.edu.cn;} and De-Cheng Zou\footnote{dczou@jxnu.edu.cn;} \\[8mm]
{College of Physics and Communication Electronics, Jiangxi Normal University, \\ Nanchang 330022, China\\[0pt] }

\vspace{40pt}
		
\underline{ABSTRACT}
\end{center}
	
Recently, an interesting \textit{doppelg\"ange} black hole solution is obtained in the string-inspired Euler-Heisenberg theory, where the black holes have the same radii but share different charges. 
We found, however, they possess different ISCOs and photon spheres, and hence affect their shadow images. In this work, we investigate the optical appearances, illuminated by an optically and geometrically thin disk, are investigated, of such black hole. One finds that doppelg\"ange black holes have different optical appearances. Even the horizon radii are the same, the size of shadows are not equal. Furthermore, we found that the large magnetic charge $Q_m$ black holes give rise to novel shadow images that the usual bright rings inside shadow are not clear,
The optical appearances illuminated by spherically accretions are also examined, and it can also identify two doppelg\"ange black holes.

	\thispagestyle{empty}
	
	\pagebreak
	%\voffset=0pt
	%\setcounter{page}{1}
	%\tableofcontents
\addtocontents{toc}{\protect\setcounter{tocdepth}{2}}
	
	%%%%%%%%%%%%%%%%%%%%%%%%%%%%%%%%%%%%%%%%

\section{Introduction}

In recent years, the study of shadow images has gained prominence due to astronomical observations of black holes. Notably, the optical appearances of the M87* and SgrA* systems obtained by the Event Horizon Collaboration in 2019 \cite{EventHorizonTelescope:2019dse} and 2022\cite{EventHorizonTelescope:2022wkp} have been significant. There are still some conjectures about their true nature—whether they are black holes, wormholes, or other compact objects. While waiting for further observations, theoretical studies on related topics have proliferated to identify these systems. For instance, there are many works on calculating shadows and weak/strong gravitational lensing effects for wormholes\cite{Liu:2022lfb,Gao:2022cds,TejeiroS:2005ltc,Nandi:2006ds,Tsukamoto:2012xs,Bronnikov:2018nub,Abe:2010ap,Schee:2021pdt,Guerrero:2021pxt,Rahaman:2021web,Peng:2021osd,Delijski:2022jjj,Huang:2023yqd,Ishkaeva:2023xny,Bouhmadi-Lopez:2021zwt}, boson stars \cite{Rosa:2022toh,Rosa:2022tfv,Rosa:2023qcv}, and naked singularities \cite{Shaikh:2018lcc,Deliyski:2023gik,Deliyski:2024wmt,Huang:2024bbs,Virbhadra:1998dy}. A widely recognized view is that these objects are black holes, although there are many black hole candidates from various theories. Thus, studies of black hole shadow images are very important.

Former investigations on black hole shadows can be traced back to Bardeen in 1972, who demonstrated that the shadow shape of a Kerr black hole is not a circle \cite{Bardeen:1973tla}. Luminet was the first to establish the optical appearance of Schwarzschild black holes surrounded by a shining and rotating accretion disk using numerical methods\cite{Luminet:1979nyg}. Gralla et al. examined the shadow images of Schwarzschild black holes illuminated by a geometrically and optically thin accretion disk using ray-tracing methods in Ref.\cite{Gralla:2019xty}. A comprehensive review on calculating shadow images is provided in Ref.\cite{Perlick:2021aok}. Furthermore, recent works on the shadow images and optical appearances of various black holes can be found in Ref.\cite{Virbhadra:2008ws,Bambi:2012tg,Cunha:2015yba,Bambi:2019tjh,Zhang:2024lsf,Peng:2020wun,Guo:2020zmf,Konoplya:2020bxa,Gyulchev:2021dvt,Zhang:2023okw,Gao:2023mjb,Falcke:1999pj,Jaroszynski:1997bw,Uniyal:2022vdu,Uniyal:2023inx,Zare:2024dtf,Chowdhuri:2020ipb,Vagnozzi:2022moj}. All these works indicate that as a major topic in gravity, it has garnered widespread attention in the scientific community.

On the other hand, an abundance of astronomical data suggests that General Relativity (GR) has its boundaries, particularly in explaining phenomena such as Dark Energy and the inflationary period of the cosmos. As a result, the reliability of GR is likely to face questions in exceptional circumstances. It is widely recognized that GR is predominantly a low-energy theory. These observations prompt us to investigate alternative gravitational theories~\cite{Boulware:1985wk,Clifton:2011jh}. 
A focal point of interest lies in the derivation of an effective four-dimensional theory, offering insights into quantum corrections
that modify Einstein’s theory of gravity. These corrections encompasses the Gauss-Bonnet term \cite{Gross:1986mw}-\cite{Kanti:1995vq} and nonlinear electromagnetic corrections~\cite{Liu:2008kt,Anninos:2008sj}.
In fact, the nonlinear electrodynamics plays an important role in these strong field regimes, where the intensity
of electromagnetic fields can become comparable to the strength of gravitational fields~\cite{Sorokin:2021tge,Sarkar:2022jgn,Allahyari:2019jqz}. During the primordial stages of the universe, for instance, when energy densities were exceptionally high, the gravitational-electromagnetic interplay was pronounced. Nonlinear electrodynamics could be also instrumental in simulating the behavior of these fields throughout cosmic evolution~\cite{Ovgun:2017iwg}.

In this work, we focus on a string-inspired theory that involves classical Euler-Heisenberg electrodynamics coupled to a non-trivial dilaton field. Very recently, Bakopoulos et al. present an exact spherically symmetric string-inspired Euler-Heisenberg black hole solution~\cite{Bakopoulos:2024hah}. Later, A. Vachher et al. examine and compare the gravitational lensing, in the strong field limit, for this magnetically charged black holes~\cite{Vachher:2024fxs}. In addition,
testing gravitational theories by the shadow images is one of the most interesting directions. Black hole system, as a strong gravity regime, is a useful tool to test the General Relativity and other theories. A well-studied theory is the GR coupled to the Euler-Heisenberg theory. 

Throughout this paper, we adopt the Planck units $8\pi G=c=1$ and the $(-,+,+,+)$ convention.

\section{The theory and the solution}

We'd like to introduce the String-inspired Euler-Heisenberg theory briefly. The action of this theory reads as~\cite{Bakopoulos:2024hah}
\begin{eqnarray}\label{action}
S=\frac{1}{16\pi}\int d^4 x\sqrt{-g}\left[R-2\partial^\mu\phi\partial_\mu\phi-e^{-2\phi}F^2-f(\phi)(2\alpha F^{\rho}_{~\sigma} F^{\sigma}_{~\lambda} F^{\lambda}_{~\gamma} F^{\gamma}_{~\rho}-\beta F^4)\right],
\end{eqnarray}
where $R$ is Ricci scalar,  $F^2=F_{\mu\nu}F^{\mu\nu}\sim E^2-B^2$ is the usual Faraday scalar, and $F^4=F_{\mu\nu}F^{\mu\nu}F_{\rho\sigma}F^{\rho\sigma}$ with usual field strength $F_{\mu\nu}=\partial_{\mu}A_{\nu}-\partial_{\nu}A_{\mu}$. Moreover, parameters $\alpha$ and $\beta$ denote the coupling constants between the scalar field $\phi$ and non-linear Euler-Heisenberg electrodynamic terms with dimensions (length)$^2$. Then, the field equations from Eq.\eqref{action} are given by
\begin{eqnarray}
&&G_{\mu\nu}-2\partial_{\mu}\phi\partial_{\nu}\phi+g_{\mu\nu}\partial^{\lambda}\phi\partial_{\lambda}\phi-2^{-2\phi}(F_{\mu}^{~\lambda}F_{\nu\lambda}-\frac{1}{4}g_{\mu\nu}F^2)\nonumber\\
&&-f(\phi)\left(8\alpha F_{\mu}^{~\rho}F_{\nu}^{~\lambda}F_{\rho}^{~\eta}F_{\lambda\eta}-\alpha  F^{\rho}_{~\sigma} F^{\sigma}_{~\lambda} F^{\lambda}_{~\gamma} F^{\gamma}_{~\delta}-4\beta F_{\mu}^{~\eta}F_{\nu\eta}F^2+\frac{1}{2}g_{\mu\nu}\beta F^4\right)=0,\label{greq}\\
&&4\Box\phi+2e^{-2\phi}F^2-\frac{df(\phi)}{d\phi}\left(2\alpha  F^{\rho}_{~\sigma} F^{\sigma}_{~\lambda} F^{\lambda}_{~\gamma} F^{\gamma}_{~\delta}-\beta F^4\right)=0,\label{KGeq}\\
&&\partial_{\mu}\Big[\sqrt{-g}\left(4F^{\mu\nu}(2\beta f(\phi)F^2-e^{-2\phi})-16\alpha F^{\mu}_{~\kappa}F^{\kappa}_{~\lambda}F^{\nu\lambda}\right)\Big]=0.\label{Maxwelleq}
\end{eqnarray}

In this paper, we assume that the scalar function $f(\phi)$ takes the following form~\cite{Bakopoulos:2024hah}  
\begin{eqnarray}\label{function}
f(\phi)=-[3\cosh(2\phi)+2]=-\frac{1}{2}\left(3e^{-2\phi}+3e^{2\phi}+4\right).
\end{eqnarray}
Considering static and spherical symmetry metric ansatz
\be\label{ds}
ds^2=-h(r)dt^2+f^{-1}(r)dr^2+r^2d\Omega^2_2,
\ee
in which $h(r)$ and $f(r)$ are functions of $r$ and $d\Omega^2_2=d\theta^2+\sin^2 \theta d\varphi^2$ the  unit $2$ sphere and Substituting the ansatz into the field equations \eqref{greq}\eqref{KGeq}\eqref{Maxwelleq}, we can obtain the black hole solution 
\begin{eqnarray}
&&h(r)=1-\frac{4M^2}{Q^2_m+\sqrt{Q^4_m+4M^2r^2}}-\frac{2(\alpha-\beta)Q^4_m}{r^6},\nonumber\\
&&f(r)=1-\frac{(Q_m^4+4M^2r^2)^{3/2}}{4M^2 r^4}+\frac{Q_m^4}{4M^2r^2}+\frac{Q_m^6}{4M^2r^4}+\frac{Q_m^2}{r^2}-\frac{(\alpha-\beta)Q_m^4(Q_m^4 + 4 M^2 r^2)}{2M^2 r^8},\nonumber\\
&&\phi(r)=-\frac{1}{2}\ln\left(\frac{\sqrt{Q^4_m+4M^2r^2}-Q^2_m}{\sqrt{Q^4_m+4M^2r^2}+Q^2_m}\right), 
\end{eqnarray}
where the Maxwell four-vector takes the form
\begin{eqnarray}
A_\mu=(0,0,0,Q_m\cos{\theta})
\end{eqnarray}
and $Q_m$ stands for the magnetic charge carried by the black hole.

\begin{figure}[t]
    \centering
    \includegraphics[width=0.47\textwidth]{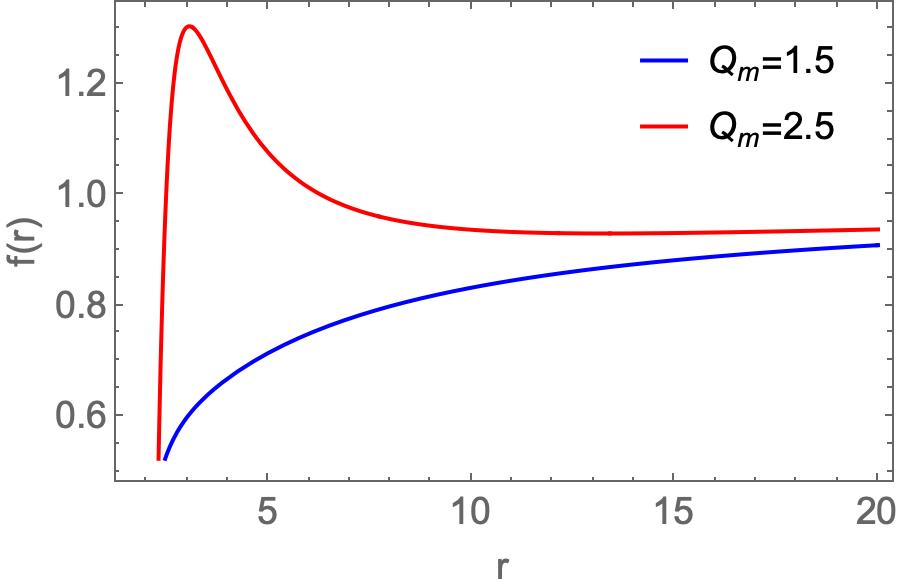}
    \caption{\textit{The metric function $g^{-1}_{rr}=f(r)$.}}
    \label{me}
\end{figure}

Considering scalar free scenario $\phi=0$, and $f(\phi=0)=1$, the black hole solution reads as
\begin{eqnarray}
h(r)=f(r)=1-\frac{2M}{r}+\frac{Q_m}{r^2}+\frac{2(\alpha-\beta)Q_m^4}{5r^6},
\end{eqnarray}
which is the Einstein-Euler-Heisenberg black hole obtained in Ref.\cite{Yajima:2000kw}.

We set $\gamma=\alpha-\beta=1$, there are two kinds of black holes, namely the single horizon with $\gamma>0$ and the double horizons with $\gamma<1$. In this work, we focus on the single horizon black holes, in which the so-called doppelg\"ange Black Holes arise. Throughout this paper, we fix $\gamma=1$ without loss of generality. The metric functions $g_{rr}^{-1}=f(r)$ are shown in Fig.\ref{me}, we found that there are two kinds of behaviors of the $f(r)$ while the $g_{tt}=h(r)$ are alway monotonically increase. For the black holes in the normal Maxwell theory, namely the RN black holes, the metric functions $f(r)$ is monotonically increasing to be $1$ in the asymptotic flat region. Interestingly, for the large $Q_m$ black holes in the string-inspired Euler-Heisenberg theory, the $g_{rr}$ is no longer monotonic. We'll see it affects the shadow images.

\section{Geodesics}

The free particles moving in a spacetime described by \eqref{ds}, there are two conserved quantities, namely the energy $E=-h(r)\dot{t}$ and the angular momentum $L=r^2\sin \theta^2 \dot{\phi}$. The dot depicts the derivative with respect to the affine parameter. Take the advantage of the spherical symmetry, we focus on the equatorial
plane, then $\sin(\theta)=1$.

It's straightforward to derive the first-order equation by the geodesic equation \cite{Huang:2023yqd},
\be\label{geoeq}
(\ft{dr}{d\varphi})^2=\ft{r^2f(r)}{h(r)}(\ft{1}{b^2}-V_{i}),
\ee
where $i=p,n$ and $b=\ft{L^2}{E^2}$ the impact parameter, and the effective potential for massive particles (time-like geodesics)
\be
V_n=\ft{h(r)}{r^2}+\ft{h(r)}{L},
\ee
while for photons (null geodesics)
\be\label{Vp}
V_p=\ft{h(r)}{r^2}.
\ee
Analysis the effective potentials yield the important orbits, such as the innermost stable circular orbit (ISCO) and photon sphere, around the black holes. Let's discuss it case by case as follows.

\subsection{Time-like geodesics and ISCO}
\label{timegeo}

As a black hole spacetime, it usually admits an special orbit to massive particles, where the particles with the specific angular momentum $L_c$ can do circular motion in the innermost orbit stably. Such orbit goes by the name of the innermost stable circular orbit (ISCO). In mathematics, the ISCO is an orbit satisfied
\be
\ft{\partial V_n}{\partial r}|_{r=r_{isco}}=0, \qquad \ft{\partial^2 V_n}{\partial r^2}|_{r=r_{isco}}=0.
\ee
One can solve an alternative equation \cite{Gao:2023mjb}
\be
r_{isco}-\frac{3h(r_{isco})h^{\prime}(r_{isco})}{2h^{\prime}(r_{isco})^2-h(r_{isco})h^{\prime\prime}(r_{isco})}=0
\ee
to obtain the ISCO.

\begin{figure}[t]
    \centering
    \includegraphics[width=0.47\textwidth]{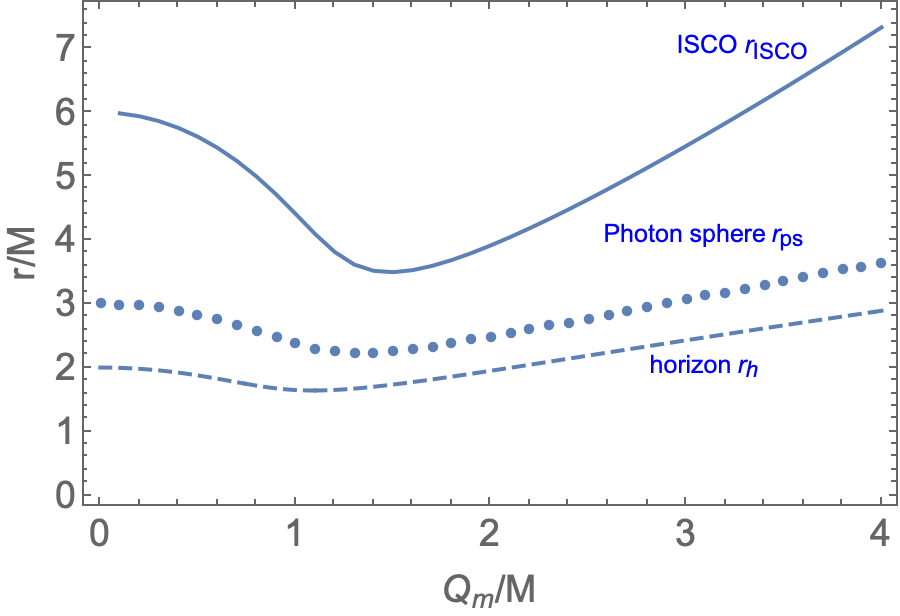}
    \caption{\textit{The radii of black hole horizon, photon sphere and ISCO as functions of magnetic charge $Q_m$.}}
    \label{Qmr}
\end{figure}

The analytical form of the ISCO is complicated and we present it as functions of $Q_m$ in Fig.\ref{Qmr}. It shows that the radii of ISCO are not monotonically increasing or decreasing with $Q_m$ increased. For the small $Q_m$ black holes, the radius of ISCO increase when the magnetic charge $Q_m$ decreased and reduce to a maximum $r_{ISCO}=6M$ when $Q_m$ vanished. Note that it coincide with ISCO of Schwarzschild black hole. For the large $Q_m$ black holes, the behavior is exactly the opposite to the small $Q_m$ case: the radius of ISCO increase along with $Q_m$ increased. Furthermore, black holes with the same mass and horizon radii but possess different $Q_m$ share not the same radii of ISCO. Hence one can distinguish them by the size of ISCOs. Moreover, the different size of ISCOs also affects the black hole shadows.

\subsection{Null geodesics and photon sphere}
\label{nullgeo}

The effective potential \eqref{Vp} has a maximum $V_{p}^{max}$ located at $r_{ps}$. This is what we called the ``photon sphere" of a black hole. The critical impact parameter $b_c=\ft{1}{V_{p}^{max}}$. According to \eqref{geoeq}, it implies that the photons with $b<b_c$ have no turning point and  will fall into the black hole horizon. An typical example of the effective potential for photons is presented in the left figure of Fig.\ref{Vp}.

\begin{figure}[h]
    \centering
\includegraphics[width=0.47\textwidth]
{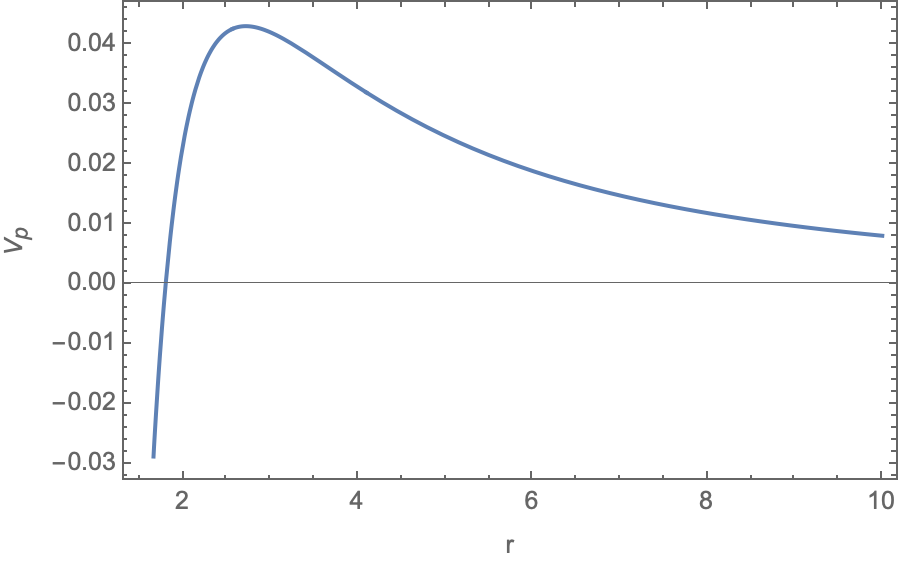}
\includegraphics[width=0.47\textwidth]{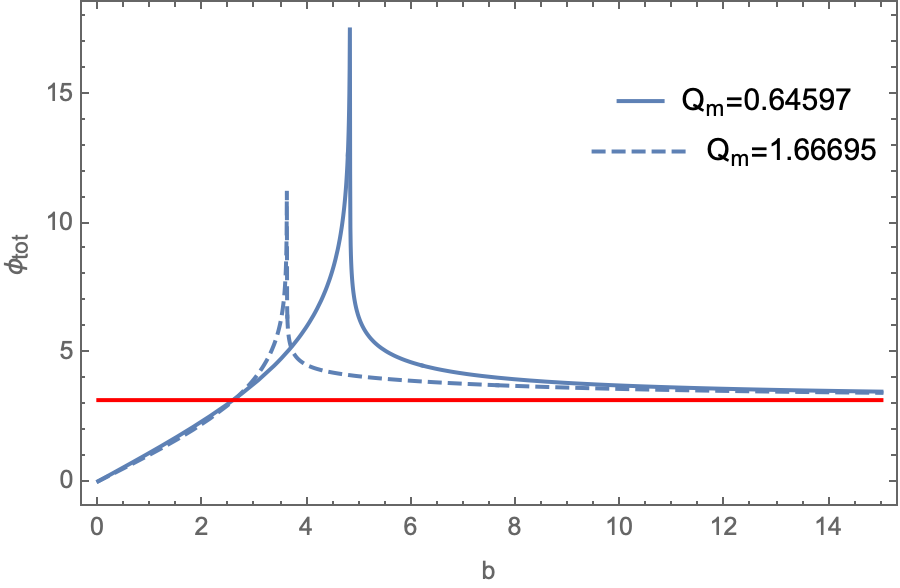}
\caption{\textit{Left figure: An typical shape of $V_p$ with $Q_m=0.64597$. Right figure: The total deflection angle of a pair of the doppelg\"ange black holes. They possess the same horizon radii and black hole masses.}}
    \label{Vp}
\end{figure}

\begin{figure}[h]
    \centering
\includegraphics[width=0.47\textwidth]{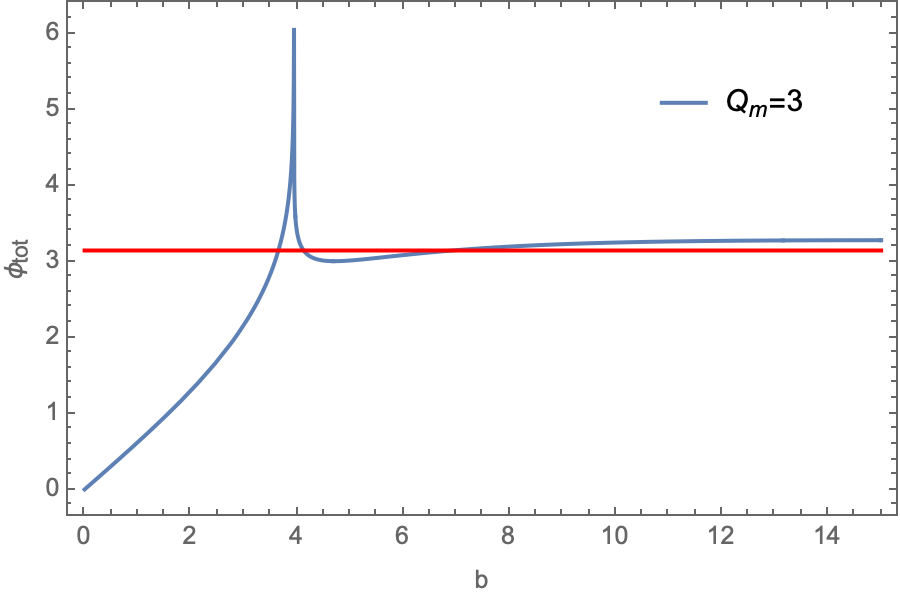}
    \caption{\textit{The total deflection angle with $Q_m=3$.}}
    \label{phitots}
\end{figure}

The photon sphere is very important in the optical appearances of black holes. We present the location of photon sphere with increasing $Q_m$ in Fig.\ref{Qmr}. Similar to the case of ISCO above, the behaviors of the photon sphere to the small and large $Q_m$ are different, and the two doppelg\"ange black holes share two different photon spheres. 

The total deflection angle is obtaining by integrating \eqref{geoeq} directly, we show the results in the right figure of Fig.\ref{Vp} and Fig.\ref{phitots}. In the thin disk model\footnote{We discuss this model in details in Sec.\ref{diskm}.}, setting $n=\ft{\phi_{tot}}{2\pi}$, one can see the three special bright regions\cite{Gralla:2019xty}:

\begin{itemize}
    \item Direct emission: created by the photons with $n<3/4$. The photons intersect with the thin disk once;
    \item Lensing ring: created by the photons with $3/4<n<5/4$. The photons intersect with the thin disk twice;
    \item Photon ring: created by the photons with $n>5/4$. The photons intersect with the thin disk more than twice;
\end{itemize}

In the right figure of Fig.\ref{Vp}, it shows that the doppelg\"ange black holes have different direct emissions, photon rings and lensing rings. In Fig.\ref{phitots}, we found a novel phenomenon on the deflection angle curve: the deflection angle is not monotonic decreasing when $b>b_c$. This only arise in the large $Q_m$ case due to the non-monotonic behavior of the metric function $f(r)$ (see Fig.\ref{me}).

\begin{figure}[t]
    \centering
\includegraphics[width=0.47\textwidth]{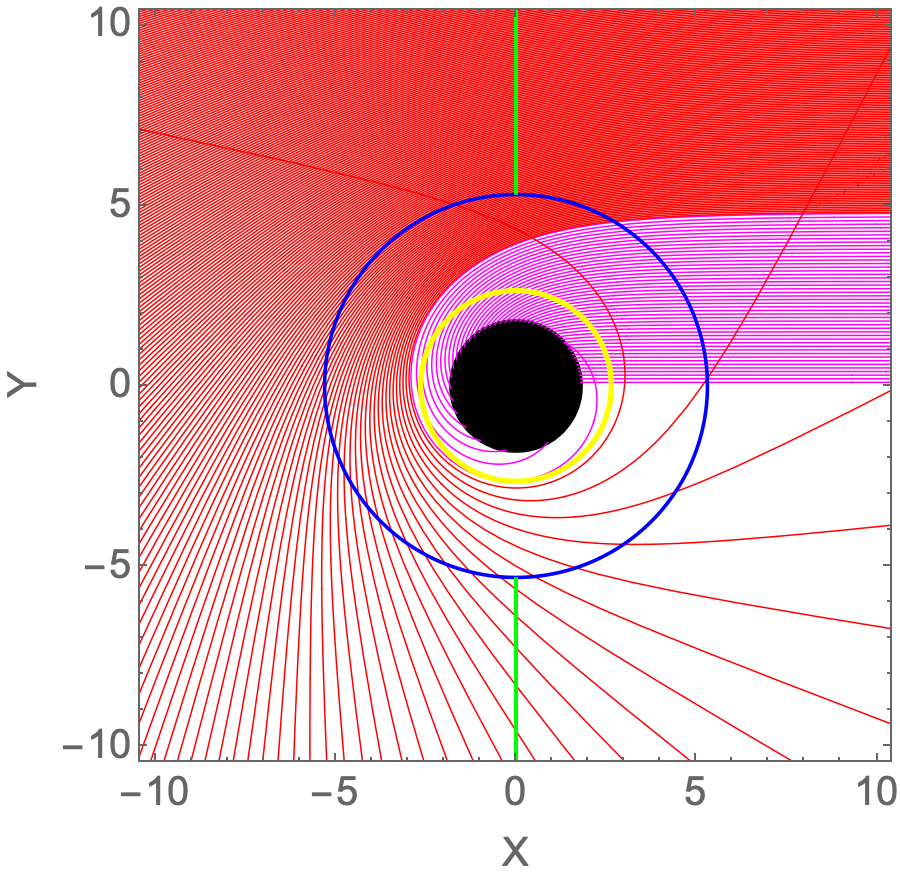}
\includegraphics[width=0.47\textwidth]{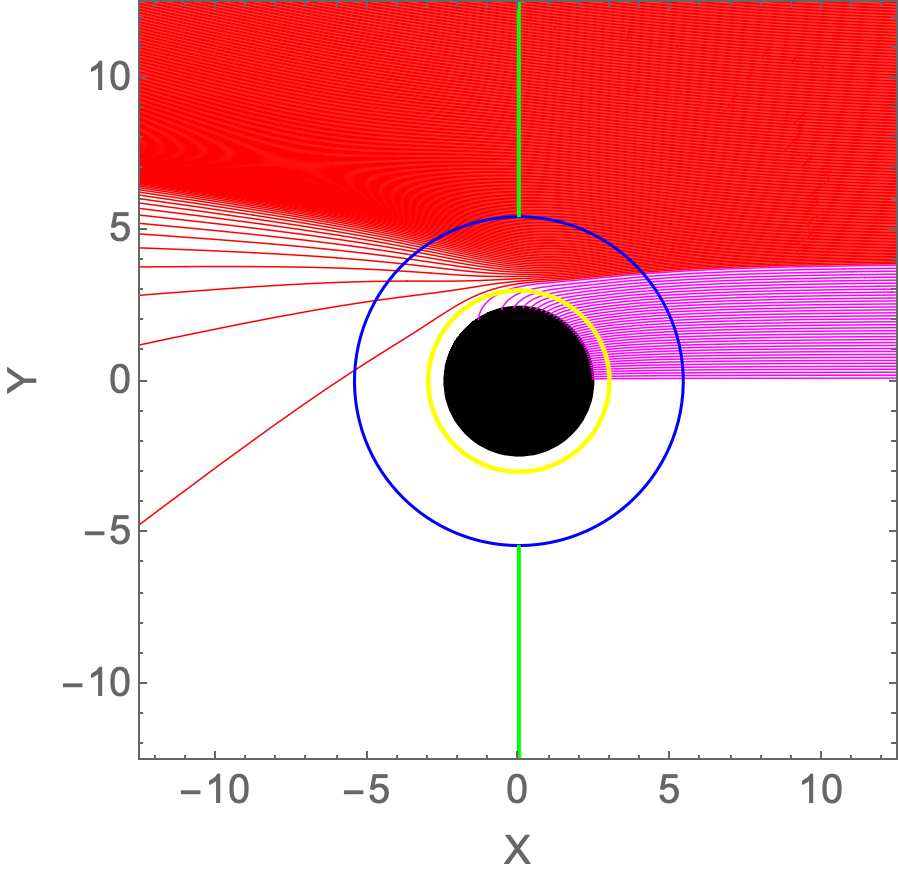}
\caption{\textit{The light rays with $Q_m=0.64597$ (right) and $Q_m=3$ (left), respectively. The red lines are the light rays which do not fall into the horizon. The purple lines are the light rays which enter the horizon. The yellow circle represents the light ring. The blue circle represents the ISCO. The green lines denote the accretion disks.}}
    \label{geo}
\end{figure}

The first-order equation \eqref{geoeq} can be solved numerically. We plot two typical figures of the geodesics in Fig.\ref{geo}. The up one in Fig.\ref{geo} is the black holes corresponding to the right figure of Fig.\ref{Vp}, where the deflection angle is monotonic decreasing when $b>b_c$. The down one in Fig.\ref{geo} corresponding the black holes in Fig.\ref{phitots}, where the repulsive effect happens in $b>b_c$ is showed.

\section{Shadow images}

Depends on the forms of accretion, it usually has two kinds of optical appearances to black holes. The fist one is the thin disk model, where the accretion disk extends from the ISCO and its intensity is monotonically decreasing. The second model is considering a spherical accretion enclosed the black hole, and it can be static or free falling to the black hole.

\subsection{Thin disk model}\label{diskm}

We consider an optically and geometrically thin disk extends from the ISCO as the light source. The shadow images are affected by intensity function on the disk. Based on the astrophysics scenarios\cite{Rosa:2022tfv}, we adopt the following models for the emitted intensity, 
\be
I_{em}(r)=I_0 (\ft{r_{ISCO}}{r})^4\ft{1+\tanh(50(r-r_{ISCO}))}{2},
\ee
where $I_0$ is a normalization factor. The ray-tracing method \cite{Falcke:1999pj,Zhang:2023okw} is used in constructing the shadow images, where the observed intensity can be derived as 
\be\label{inten}
I_{obs}(b)=
\sum_n g_{tt}^2\, I_{em}(r)|_{r=r_n(b)},
\ee
where $r_n(b)$ is the transfer function that indicates the radii of the $n$th intersection with the accretion disk. 

\begin{figure}[h]
    \centering
\includegraphics[width=0.40\textwidth]{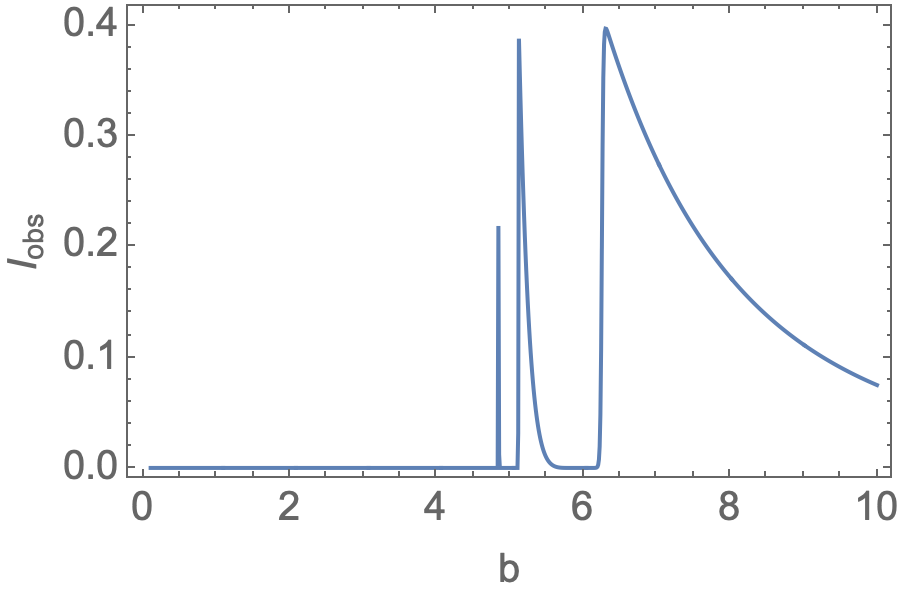}
\includegraphics[width=0.40\textwidth]{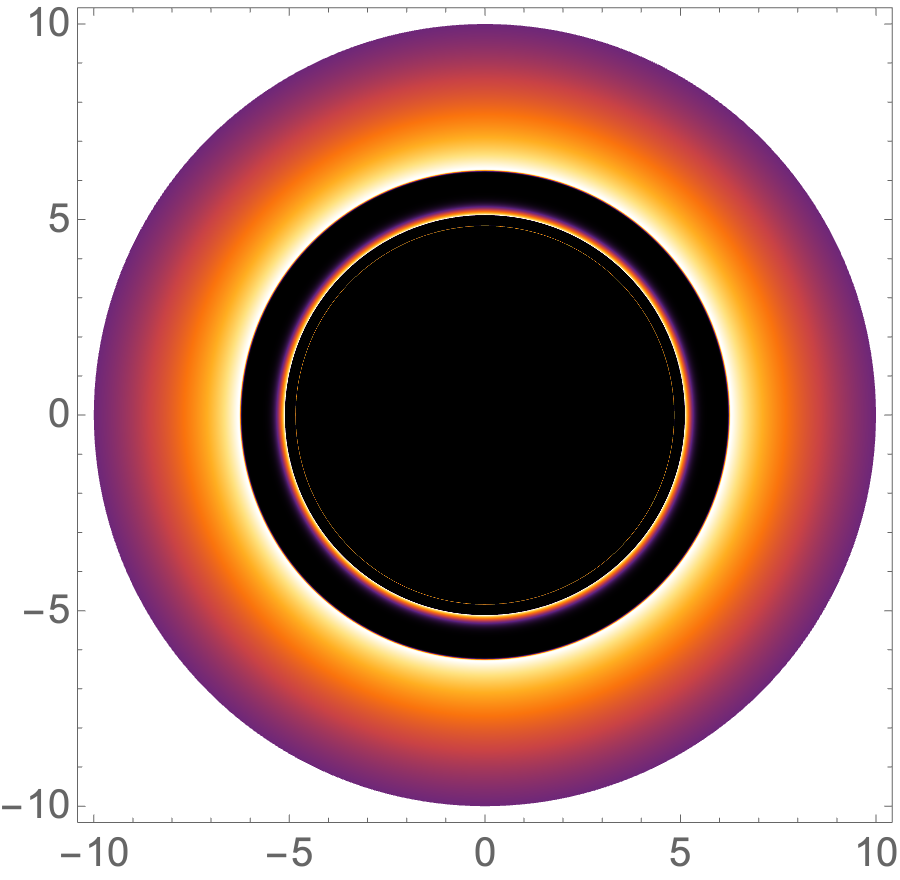}
    \caption{\textit{The observed intensity and the corresponding shadow image with $Q_m=0.64597$.}}
    \label{s1}
\end{figure}

\begin{figure}[h]
    \centering
\includegraphics[width=0.40\textwidth]{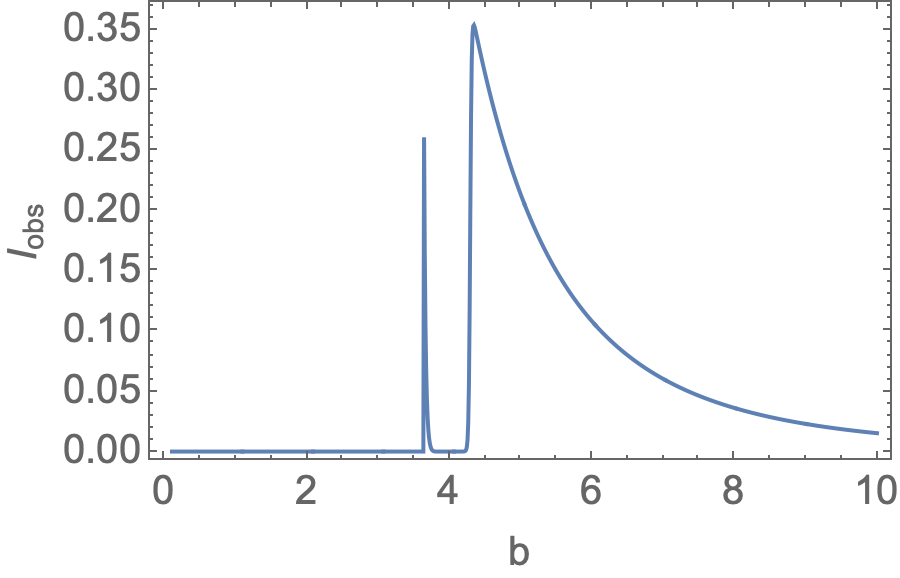}
\includegraphics[width=0.40\textwidth]{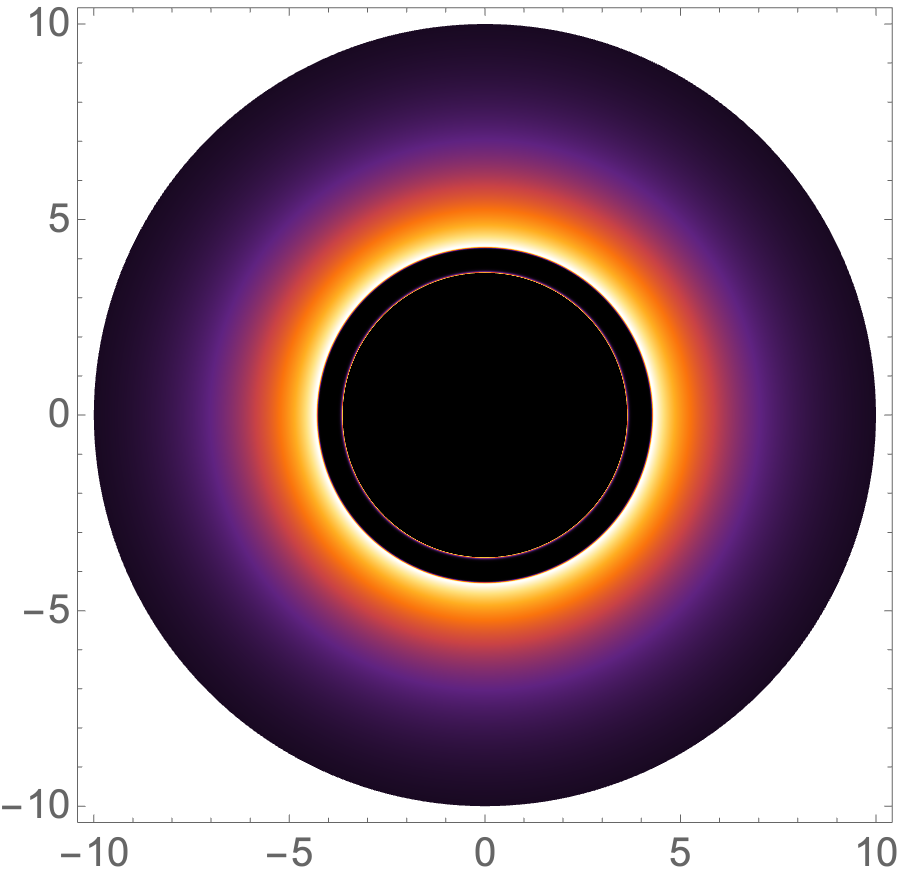}
    \caption{\textit{The observed intensity and the corresponding shadow image with $Q_m=1.6669$.}}
    \label{s2}
\end{figure}

\begin{figure}[h]
    \centering
\includegraphics[width=0.40\textwidth]{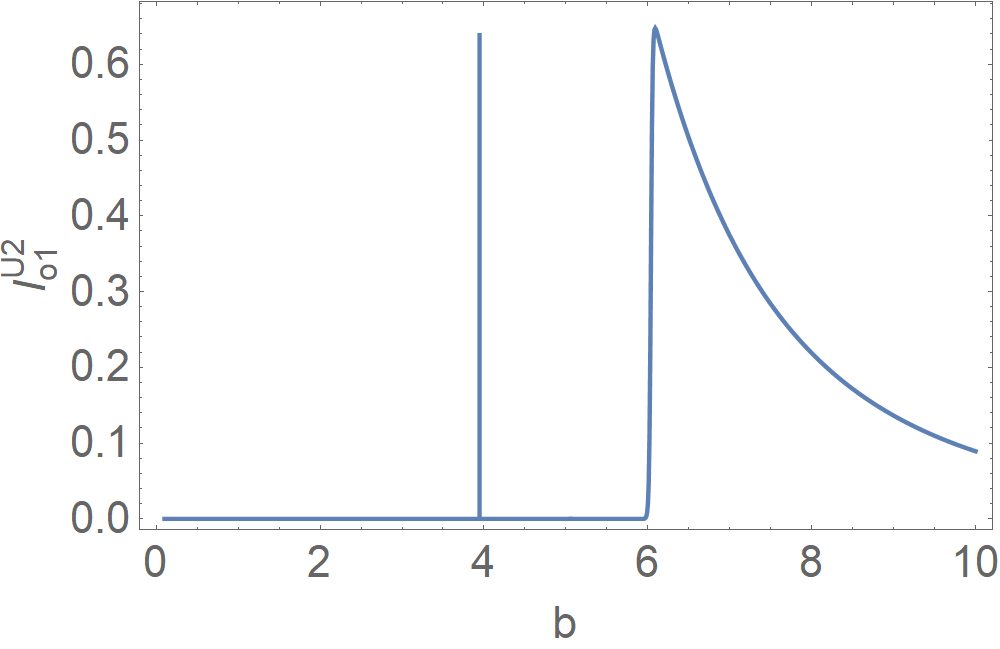}
\includegraphics[width=0.40\textwidth]{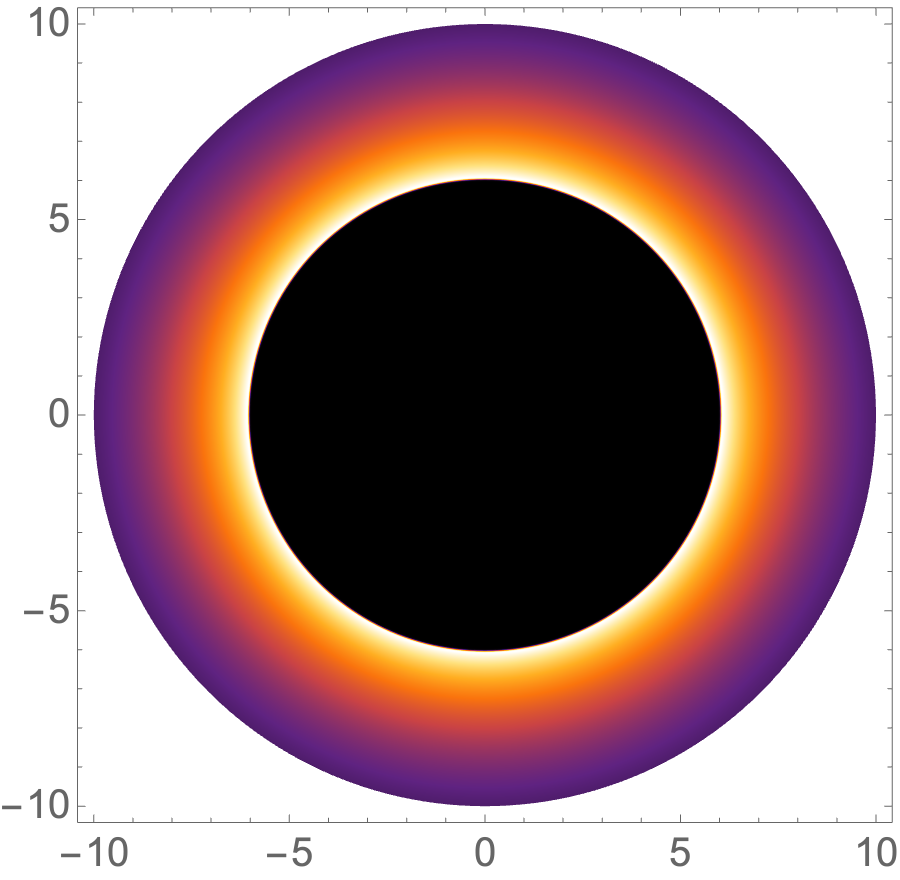}
\caption{\textit{The observed intensity and the corresponding shadow image with $Q_m=3$. The width of the peak around $b=4$ of the observed intensity is about $\Delta b\sim 10^{-4}$, such that it is hard to be observed.}}
    \label{s3}
\end{figure}

The numeric results of the observed intensity and the corresponding shadow images to the doppelg\"ange black holes are shown in Fig.\ref{s1} and Fig.\ref{s2}. It shows that the doppelg\"ange black holes can be distinguished from the shadow images illuminated by the thin accretion disk model. This is mainly due to the different radii of their photon spheres and ISCOs, even for such black holes share the same horizon radii and mass.

As we discuss above, the metric function $g_{rr}=f(r)$ are no longer monotonic increased in the large $Q_m$ case. In Fig.\ref{s3}, we present the results related to a large $Q_m$ case. We see, because of the unusual behavior of the metric function, there is no clear bright ring inside the shadow.

\subsection{Static spherical accretion model}

Another possible accretion model in astrophysics is the spherical accretion flow model. As a ideal model, it usually consider the accretion distributed outside the black hole horizon with spherically. 

\begin{figure}[h]
    \centering
\includegraphics[width=0.40\textwidth]{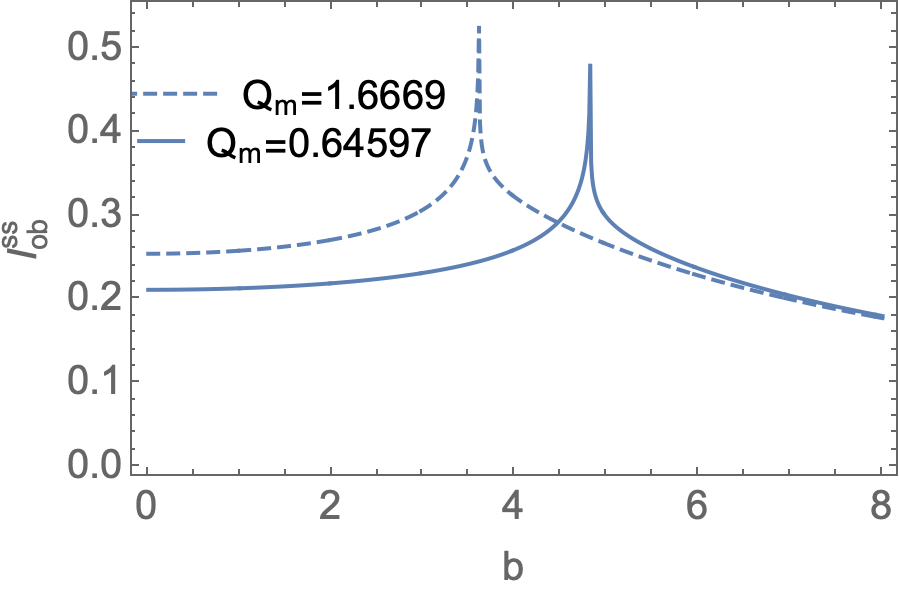}
    \caption{\textit{The observed intensity $I_{ob}^{ss}$ of a pair of the doppelg\"ange black holes with the static spherical accretion model.}}
    \label{iobs1}
\end{figure}

To begin with, we consider the static accretion model. The observed specific intensity of photon with a frequency $\nu_o$ to an distant observer is given by \cite{Jaroszynski:1997bw,Bambi:2013nla}
\be
I(\nu_o)=\int g^3 j(\nu_e)dl_{prop},
\ee
where $\nu_e$ is the intrinsic photon frequency, $g=\ft{\nu_o}{\nu_e}=\sqrt{h(r)}$ the red-shift factor, the infinitesimal proper length is denoted by $dl_{prop}$ and $j(\nu_e)$ is the emissivity per unit volume in the rest frame of the emitter. Then total observed intensity is a sum up of all the possible specific intensity,
\be\label{iobs}
I_{obs}^{ss}=\int I(\nu_o) d\nu_o.
\ee
Furthermore, we consider the radiation is monochromatic with fixed a frequency $\nu_f$, which means 
\be
j_{\nu_e}\propto \ft{\delta(\nu_e-\nu_f)}{r^2}.
\ee
Finally, the observed intensity \eqref{iobs} reduces to
\be
I_{obs}^{ss}=\int \ft{h(r)^2}{r^2}\sqrt{\ft{1}{f(r)}+\ft{h(r) b^2}{f(r)(r^2-b^2 h(r)^2)}}dr.
\ee
To the doppelg\"ange black holes, they give rise to different $I_{obs}^{ss}$. We show an example in Fig.\ref{iobs1}, and the related images are shown in Fig.\ref{shadowss}

\begin{figure}[h]
    \centering
\includegraphics[width=0.40\textwidth]{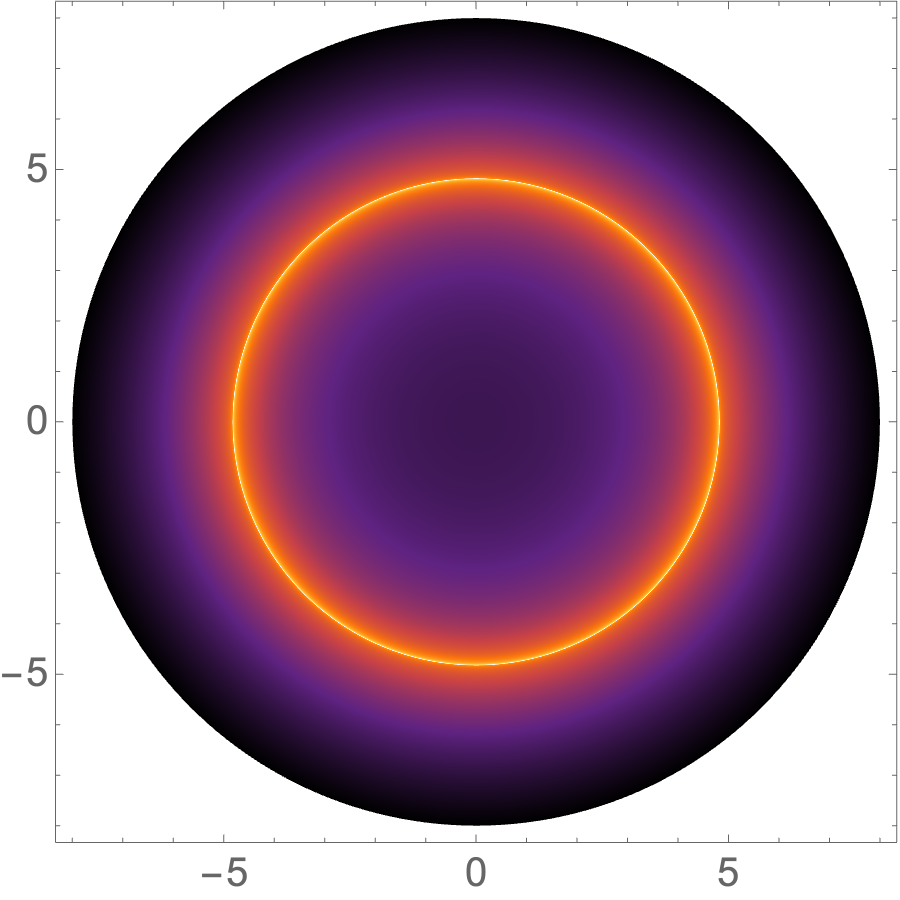}
\includegraphics[width=0.40\textwidth]{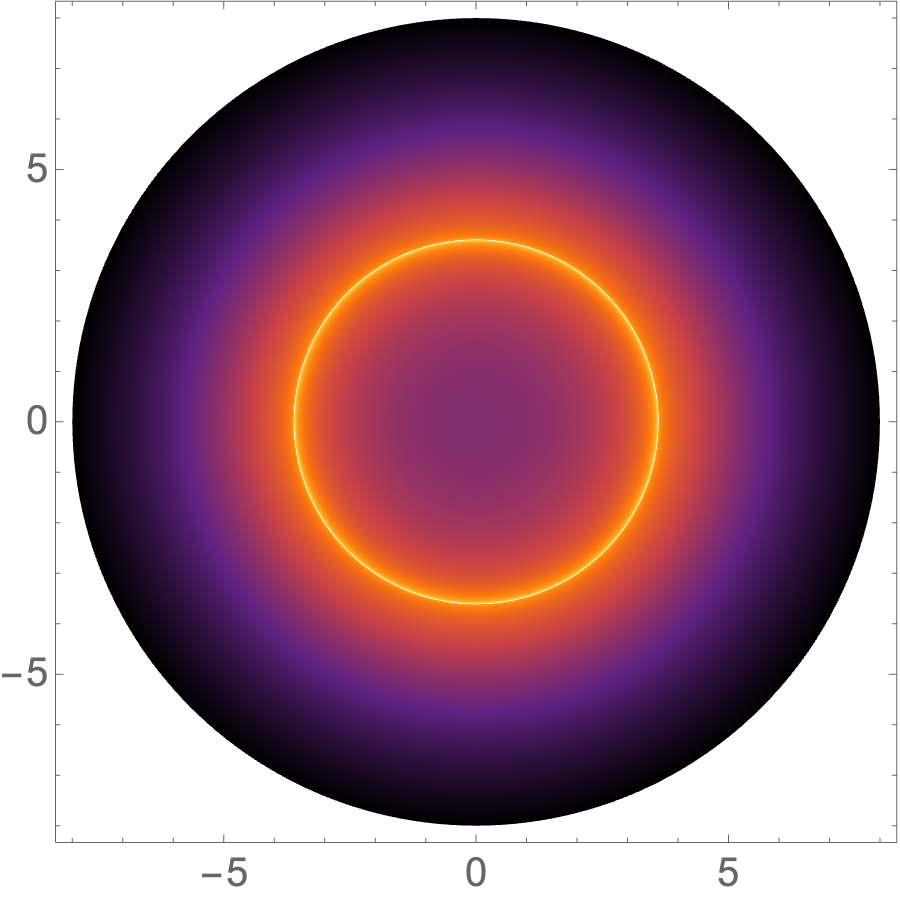}
    \caption{\textit{With the static spherical accretion model, the shadow images of the black holes with $Q_m=0.64597$ and $Q_m=1.6669$, respectively.}}
    \label{shadowss}
\end{figure}

The large $Q_m$ black holes do not reveal novel features in the static model, hence we don't show them here. However, the more realistic case is that the matter falls into the horizon, namely the the infalling spherical accretion flow model. Then the large $Q_m$ black holes are quite different from the small $Q_m$ one.

\subsection{Infalling spherical accretion flow model
}

\begin{figure}[h]
    \centering
\includegraphics[width=0.47\textwidth]{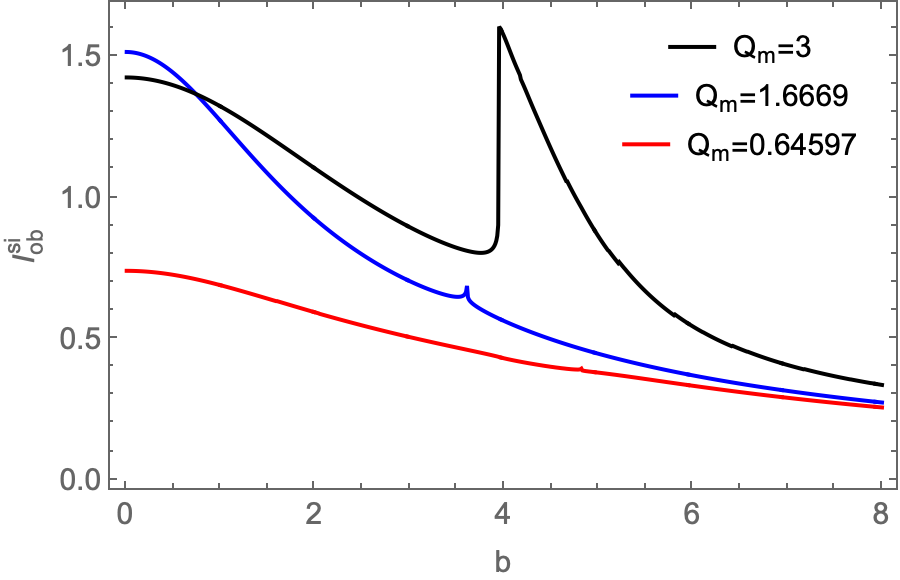}
\caption{\textit{The observed intensity $I_{ob}^{si}$ with the infalling spherical accretion flow model.}}
    \label{iobs2}
\end{figure}

The \eqref{iobs} is hold for infalling accretion case. But the red-shift factor depends the velocity of the accretion flow, namely,
\be
g=\ft{k_\mu u^\mu_o}{k_\nu u^\nu_e},
\ee
where $(k^\mu, u^\mu_o, u^\nu_e)$ are the four-velocities of the photons, the stationary distant observers and the infaling accretion flow, respectively. It is straightforward to obtain all the non-vanished components of all the four-velocities are
\be
k_t=\ft{1}{b}, \qquad k_r=\sqrt{\ft{1}{b^2h(r)f(r)}-\ft{1}{f(r)r^2}},\qquad u^t_e=\ft{1}{h(r)}, \qquad u^r_e=-\sqrt{\ft{f(r)}{h(r)}-f(r)},
\ee
and $u^\mu_o=(1,0,0,0)$. Finally, the observed intensity of the infalling accretion flow model is given by
\be
I_{obs}^{si}=\int \ft{g^3}{r^2}\bigg(\sqrt{\ft{1}{h(r)}(\ft{1}{f(r)}-\ft{b^2 h(r)}{f(r)r^2})}\bigg)^{-1}dr.
\ee

\begin{figure}[ht]
    \centering
\includegraphics[width=0.40\textwidth]{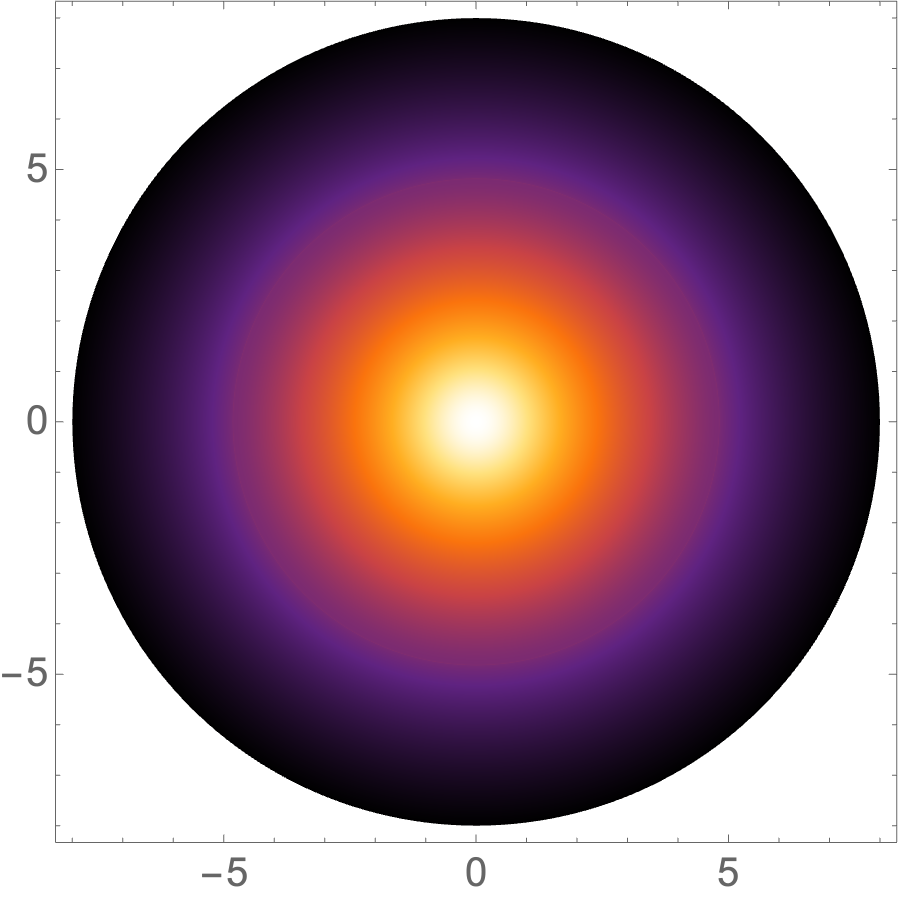}
\includegraphics[width=0.40\textwidth]{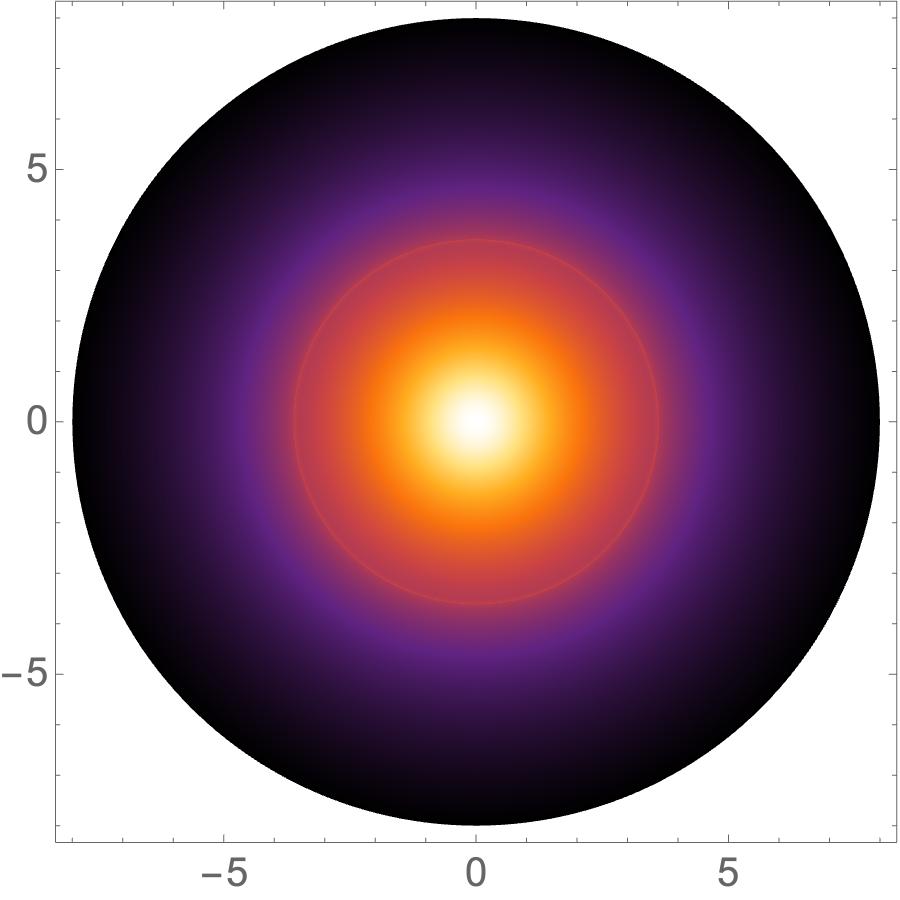}
\caption{\textit{With the infalling spherical flow accretion, the shadow images of black holes with $Q_m=0.64597$, $Q_m=1.6669$ respectively.}}
    \label{shadowss}
\end{figure}

\begin{figure}[ht]
    \centering
\includegraphics[width=0.40\textwidth]{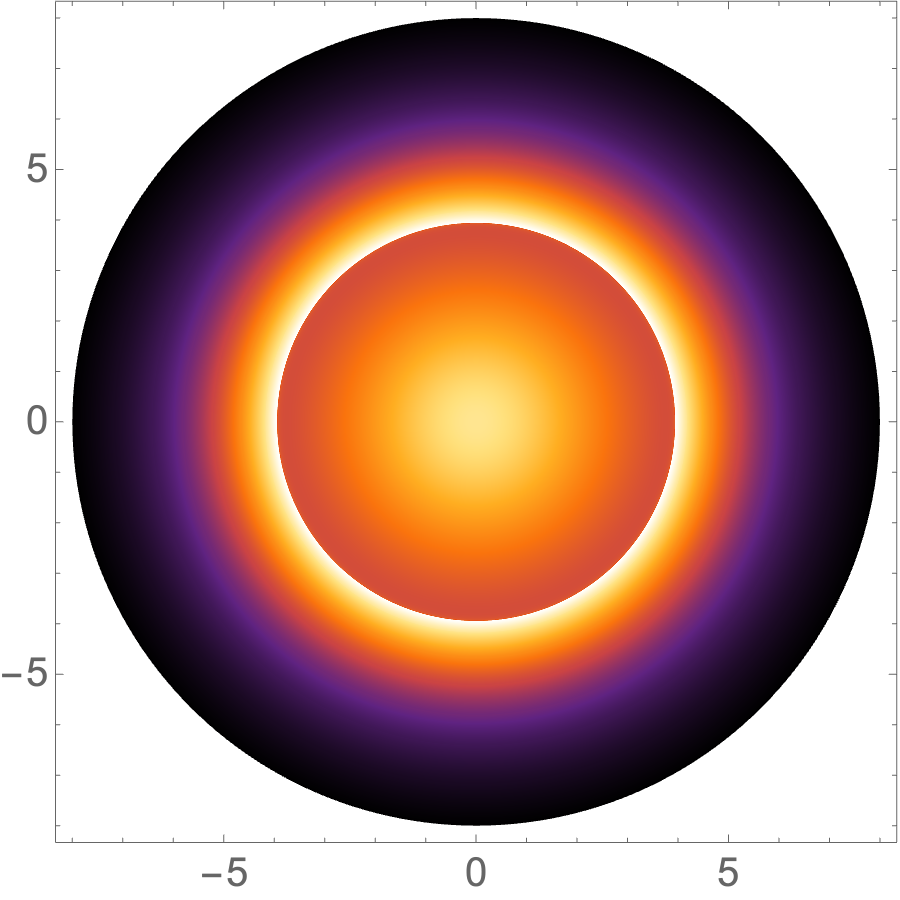}
\caption{\textit{With the infalling spherical flow accretion, the shadow images of black holes with $Q_m=3$.}}
    \label{shadowss2}
\end{figure}

Three examples of the $I_{obs}^{si}$ corresponding to the doppelg\"ange black holes and large $Q_m$ black hole are shown in Fig.\ref{iobs2}. It it clear that the we could also distinguish the doppelg\"ange black holes in this model. Furthermore, the $I_{obs}^{si}$ of the large $Q_m$ black hole shows a distinct peak. We present the shadow images related to these cases in Fig.\ref{shadowss}.

\section{Conclusion and Outlook}
\label{sec:conclusion}

The phenomenon of having doppelg\"ange black holes in the string-inspired Euler-Heisenberg theory is quite attractive. In Ref.\cite{Bakopoulos:2024hah}, the authors pointed out that a pair of the doppelg\"ange black holes could share different thermodynamics. In our work, we have investigated the optical appearances of such black holes in detail. The analysis of time-like and null geodesics is examined. We found even for the black holes have the same mass and horizon radii, the ISCOs and photon spheres are intrinsically different. It follows that the shadow images cast by the doppelg\"ange black holes could be also different. We have studied three types of accretion models, including the thin disk, static and infalling spherical flow accretions, to establish the shadow images. Our results reveal that one can easily identify the doppelg\"ange black holes by their optical appearances. It would be really helpful for future observations.

On the other hand, we found the black holes with large magnetic charge $Q_m$ in this theory play an intriguing property. Unlike the RN solution, the metric function $g_{rr}$ is non-monotonic with $r$. This fact leads the total deflection angle to be unusual. The repulsive effect arises in the vicinity of the photon sphere, which gives rise to an unclear bright ring inside the shadow. This would be a novel feature and can be detected in future observations.

For future works, one of the interesting directions is to explore the rotating doppelg\"ange black holes. It is a question that will the rotation affect the difference between the rotating doppelg\"ange black holes? Furthermore, astronomical black holes usually contain rotation, thus the results on the rotating doppelg\"ange black holes would be more realistic. 

One would ask, are there doppelg\"ange black hole solutions apart from the string-inspired Euler-Heisenberg theory? Ignoring the no-hair conjecture, the answer could be ``No". Then it is valuable to check the shadow properties of doppelg\"ange black holes in other theories. The results could be useful to clarify the gravitational theories.

\section*{Acknowledgment}

XYK and HH are gratefully acknowledge support by the National Natural Science Foundation of China (NSFC) Grant No.~12205123 and Jiangxi Provincial Natural Science Foundation
with Grant No.~20232BAB211029. M. Y. L is supported by the National Natural Science Foundation of China with Grant No. 12305064 and Jiangxi Provincial Natural Science Foundation with Grant No. 20224BAB211020. D. C. Z is supported by National Natural Science Foundation of China (NSFC) (Grant No. 12365009) and Jiangxi Provincial Natural Science Foundation (No. 20232BAB201039).

\end{document}